\documentclass[12pt]{spieman}  % 12pt font required by SPIE;
\usepackage{amsmath,amsfonts,amssymb}
\usepackage{graphicx}
\usepackage{setspace}
\usepackage{tocloft}

\usepackage{pstricks}

 \hypersetup{breaklinks}

\title{Kernel formalism applied to Fourier-based wave front sensing in presence of residual phases}

\author[1,2,*]{Olivier Fauvarque}
\author[2,3]{Pierre Janin-Potiron}
\author[2]{Carlos Correia}
\author[2]{Yoann Br\^ul\'e}
\author[2]{Benoit Neichel}
\author[2]{Vincent Chambouleyron}
\author[2,3]{Jean-Francois Sauvage}
\author[2,3]{Thierry Fusco}

\affil[1]{Aix-Marseille Université, CNRS, Institut Fresnel, F-13013 Marseille, France}
\affil[2]{Aix Marseille Université, CNRS, CNES, LAM (Laboratoire d'Astrophysique de Marseille) UMR 7326, 13388, Marseille, France}
\affil[3]{ONERA--the French Aerospace Laboratory, F-92322 Chatillon, France }
\affil[*]{Corresponding author: olivier.fauvarque@lam.fr}

\cftpagenumbersoff{figure}
\cftpagenumbersoff{table}

\begin{document}
\maketitle

\begin{abstract}
In this paper, we describe Fourier-based Wave Front Sensors (WFS) as linear integral operators, characterized by their Kernel. In a first part, we derive the dependency of this quantity with respect to the WFS's optical parameters: pupil geometry, filtering mask, tip/tilt modulation. In a second part we focus the study on the special case of convolutional Kernels. The assumptions required to be in such a regime are described. We then show that these convolutional kernels allow to drastically simplify the WFS's model by summarizing its behavior in a concise and comprehensive quantity called the WFS's Impulse Response. We explain in particular how it allows to compute the sensor's sensitivity with respect to the spatial frequencies. Such an approach therefore provides a fast diagnostic tool to compare and optimize Fourier-based WFSs. In a third part, we develop the impact of the residual phases on the sensor's impulse response, and show that the convolutional model remains valid. Finally, a section dedicated to the Pyramid WFS concludes this work, and illustrates how the slopes maps are easily handled by the convolutional model.
\end{abstract}

\section{Introduction}

Linear integral operators are the continuous version of matrices. They linearly transform an input into an output depending on a quantity called Kernel. Mathematically, such an operation may be written:
\begin{equation}
\text{Output}|_X = \int~\text{d}x~\textbf{K}|_{X;x}~\text{Input}|_{x} \label{fonda}
\end{equation}
$x$ (resp. $X$) is the variable of the input (resp. output) space, $\textbf{K}$ making the link between those two spaces.  Examples of integral operators are many: Fourier, Hilbert, Laplace transforms, etc. are the most famous. In a more general framework, it is absolutely relevant to try to describe a continuous linear system thanks to an integral operator. Advantages of such an approach are many. First of all, the Kernel allows to synthesize the system's behavior into a compact and elegant mathematical quantity. Moreover, due to the fact linear integral operators may be seen as the continuous version of matrices (see Appendix \ref{tata}), the use of Kernel formalism may improve the understanding of the discrete description of systems that numerical approach often requires. We finally note that reconstructing the input from the output, i.e. inverting   \eqref{fonda}, depends on our capability to find the "inverse Kernel" $\textbf{K}^{-1}$ of the system. 
Fortunately, depending on the nature of $\textbf{K}$, many methods already exist to calculate this inverse if it does exist. Knowing $\textbf{K}$ is thus critical to build robust and relevant reconstruction algorithms. 

The purpose of this article is to use this powerful formalism to study optical systems which probe the wavefront of the light. They are called Fourier-based Wave Front Sensors \cite{Fauvarque16} and are essentially used in Adaptive Optics for Astronomy. The first part of this paper provides an optical description of Fourier-based WFSs. We make explicit the input and output of such sensors and introduce proper mathematical quantities to characterize the different optical elements.  In a second part, we give the dependency of the Kernel with respect to these optical elements. The third part is dedicated to the particular case of convolutional Kernel. Phase reconstruction, in that situation, is greatly facilitated since the inverse Kernel has an explicit formulation. Moreover, it allows to define the WFS's Impulse Response and Transfer Function which are compact and meaningful quantities allowing, for instance to rapidly visualize the WFS's sensitivity (see Appendix \ref{ujuj}). Since the convolutional Kernel case is not systematic, we give the assumptions it relies on. Another part is dedicated to generalization of these results in presence of static and dynamic residual as it is often the case, in practice, in Adaptive Optics. These results are finally applied to the most used Fourier-based WFS which is the Pyramid WFSensor \cite{Ragazzoni1996} in Appendix \ref{PyrPyr}.

\section{Fourier-based Wave Front Sensing}

\subsection{Optical system}

\begin{figure}[htbp]
\centering
\includegraphics[scale=0.30]{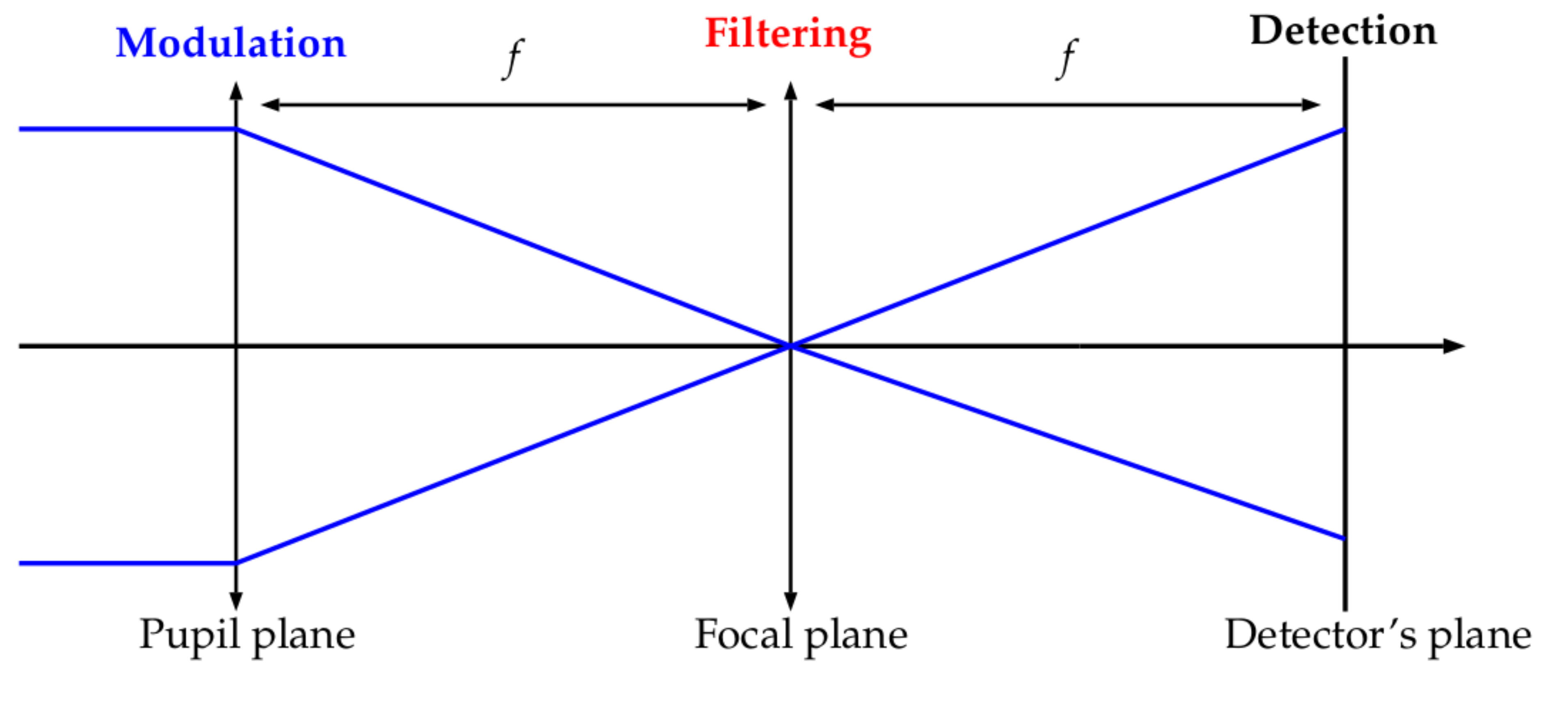}
\caption{Schematic view (in 1D) of a Fourier filtering optical system\label{f1}}
\end{figure}

We consider the optical system shown in Fig. \ref{f1}. The first plane corresponds to the pupil plane of the telescope. It contains a focusing device which is described as a perfect lens with a focal $f$, an aperture and another element that we call modulation. We describe it in detail in the next paragraph. The second plane is the focal plane which contains a filtering mask mathematically described by its transparency function $m$ and an imaging lens with focal $f/2$.  The detector is finally placed in the next plane which is conjugated to the first pupil plane. 

\subsection{Optical propagation}

The incoming field is called $\psi_i$, we assume that its light is monochromatic at wavelength $\lambda$.
In the Adaptive Optics Wave Front Sensing context, the field is windowed by an entrance pupil. We call its phase $\phi$ for turbulent phase and describe the pupil geometry thanks to its indicator function $\mathbb{I}_P$. Besides, we assume that the total flux is unitary. Under these assumptions, we have:
\begin{equation}
\psi_i = \mathbb{I}_P e^{\imath \phi}
\end{equation}
Let's note that turbulent phase has \emph{a priori} an infinite support. Nevertheless, due to windowing by the finite aperture, only the part which goes through the entrance pupil has a physical interest and may be coded by the WFS. As a consequence, the quantity to be measured, i.e. the sensor's input is the turbulent phase windowed by the pupil:
\begin{equation}
\text{WFS's input:}~~~~~\mathbb{I}_P\phi
\end{equation}

The incoming field is then shaped by the "modulation". Modulation consists in adding a dynamic aberration to the field via a controlled moving device placed in the entry pupil plane. The movement is regular and may be described as a cycle (although it can be generalised). The detector is then synchronized with it in order to have one image per cycle. This system was initially introduced by \cite{Ragazzoni1996} in order to adjust the Pyramid WFS's linearity range. It is specific to this sensor yet it should remain general. In this paper, we are interested in the particular case of "tip/tilt modulation" which consists in adding a tip/tilt aberration in the field. This modulation phase is called $\phi_\text{mod}$ and defined as:
\begin{equation}
\phi_\text{mod}(\vec{\boldsymbol{\alpha}})|_{\vec{\textbf{r}}} = \frac{2\pi}{\lambda}\vec{\boldsymbol{\alpha}}.\vec{\textbf{r}}\label{ttm}
\end{equation}
The vector $\vec{\textbf{r}}$ codes the position in the pupil plane while $\vec{\boldsymbol{\alpha}}$ codes the amplitudes of the modulation tip and tilt. Brief clarifications about notations: $\phi_\text{mod}(\vec{\boldsymbol{\alpha}})|_{\vec{\textbf{r}}}$ means that the \emph{function} $\phi_\text{mod}$ depends on one \emph{true spatial} vector $\vec{\textbf{r}}$ but also on another \emph{non-spatial} parameter $\vec{\boldsymbol{\alpha}}$. 
To describe the modulation, we also need to indicate the time spent for each modulation phase. To do so, we introduce the modulation weighting function $w$. Usually, this function directly depends on the time variable but we prefer here to use as variable the tip/tilt amplitudes $\vec{\boldsymbol{\alpha}}$:
\begin{equation}
\begin{array}{ccccc}
w & : & \mathbb{R}^2 & \to & \mathbb{R}_+ \\
 & & \vec{\boldsymbol{\alpha}} & \mapsto & w|_{\vec{\boldsymbol{\alpha}}} \\
\end{array}
\end{equation}
The weighting function is now a 2D function and $w|_{\vec{\boldsymbol{\alpha}}}$ codes the time spent for the modulation phase $\phi_\text{mod}(\vec{\boldsymbol{\alpha}})$.  Such an approach presents several advantages. First, it allows to visualize the tip/tilt modulation in its \emph{natural} focal plane. Indeed, in this plane $\vec{\boldsymbol{\alpha}}$ vector correspond to a spatial shift (that's why, we do not use the notation $w(\vec{\boldsymbol{\alpha}})$ but $w|_{\vec{\boldsymbol{\alpha}}}$). Subsequently, the weighting function directly gives the profile of the modulation. Thus, it becomes possible to envision 2D modulation, as for instance a disk modulation (the classical description only allowed 1D modulation, as for instance ring modulation).
To ensure the energy conservation, we finally enforce that the weighting function has a unitary 1-norm:
\begin{equation}
\int_{\mathbb{R}^2} w|_{\vec{\boldsymbol{\alpha}}}~\text{d}^2 \vec{\boldsymbol{\alpha}}= 1
\end{equation}

Regarding the optical Fourier filtering stage, it is based on the fact that the focal plane corresponds to the reciprocal space (i.e. the spatial frequencies space) of the phase (or pupil) space. This fact is directly linked to Fraunhofer's diffraction. Subsequently, a mask placed in this plane acts like a spatial frequencies filter. A relevant way to describe this mask consists in using its transparency function $m$. For a more detailed description of this quantity, the curious reader may refer to \cite{Fauvarque16}.

Describing mathematically the filtering process consists in getting the field in the detector's plane, we call it $\psi_d(\vec{\boldsymbol{\alpha}})$ since it depends on the modulation phase. Optical propagation laws indicate that this field equals to the convolution product  between the pupil plane field and the Fourier transform of the transparency function of the mask:
\begin{equation}
\psi_d(\vec{\boldsymbol{\alpha}})=\Big[\mathbb{I}_P \exp\big(\imath (\phi+ \phi_\text{mod}(\vec{\boldsymbol{\alpha}})\big)\Big]\star \hat{m}
\end{equation} 
where $\hat{.}$ is the Fourier transform and $\star$ the 2D convolution product.
The last step consists in converting the detector's field into photo-electrons. Physically, this detection consists in the integration during the modulation cycle of the squared modulus of $\psi_d$. The resulting intensity, which depends on the phase $\phi$, equals to:
\begin{equation}
I(\phi) = \int_{\mathbb{R}^2}  \text{d}^2 \vec{\boldsymbol{\alpha}}~w|_{\vec{\boldsymbol{\alpha}}}~\\
\left| \Big[\mathbb{I}_P \exp\big(\imath (\phi+ \phi_\text{mod}(\vec{\boldsymbol{\alpha}})\big)\Big]\star \hat{m} \right|^2\label{II}
\end{equation}

In the wave front sensing context, it is relevant to give the dependence of this intensity with respect to the phase. A way to efficiently describe this dependence has been shown in detail in \cite{Fauvarque16}. It consists in doing a Taylor's development on the phase term:
\begin{equation}
\exp(\imath \phi) = \sum_{q=0}^\infty \frac{\imath^q \phi^q}{q!}
\end{equation}
which allows then to decompose the intensity into phase-constant, linear, quadratic, cubic, etc. terms: 
\begin{equation}
I(\phi) =I_{\text{constant}} + I_\text{linear}(\phi) + I_\text{quadratic}(\phi) + I_{\text{cubic}}(\phi)+...
\end{equation}
We note that the constant term does not contain any information about phase since it does not depend on $\phi$. Consequently, it is common practice to apply a \emph{return-to reference} operation on the intensity which consists in numerically subtracting $I_\text{constant}$ to $I$. The resulting quantity which may be seen as the WFS's output is called \emph{differential intensity}:
\begin{eqnarray}
\Delta I(\phi)&\equiv & I(\phi)-I_\text{constant}\label{mI}\\
&=&I_\text{linear}(\phi) + I_\text{quadratic}(\phi) + ...\label{DLL}
\end{eqnarray}
The first phase dependence of $\Delta I$ is thus the linear one. We note that this numerical operation is easy to do in practice since $I_\text{constant}$ equals to the intensity when there is no turbulent phase, i.e. $I(0)$. Finally, we specify that this return-to-reference operation is in practice associated to a flux normalization which is not necessary here since the flux has been assumed unitary. 

\section{Linear Model}

A standard approximation consists in assuming that the WFS works in its linearity regime. Mathematically, it means that the differential intensity only contains the linear term:
\begin{equation}
\Delta I(\phi) \approx I_{\text{linear}}(\phi)
\end{equation}
Calculating the linear regime of a WFS, i.e. the phase subspace where non-linear terms may be neglected, is a challenging task which requires the study of the WFS's dynamic range but it is not the topic of this paper. Subsequently, we consider that the linear regime of a WFS essentially corresponds to the small phases domain:
\begin{equation}
\phi << 1
\end{equation}
Such an assumption does make sense in Adaptive Optics context since most of the wave front sensing is done in closed loop, i.e. when phase-to-be-measured are residuals of the atmospheric turbulent phase.

\subsection{Linear intensity}\label{fffr}

Within the small phases approximation framework, WFS's output equals the linear intensity.  We give an explicit expression of it initialy developed in \cite{Fauvarque16}:
\begin{equation}
I_\text{linear}(\textcolor{black}{\phi})=2 \textbf{Im}\bigg[\int_{\mathbb{R}^2} \text{d}^2 \vec{\boldsymbol{\alpha}}~~w|_{\vec{\boldsymbol{\alpha}}}~ \Big(\big(\mathbb{I}_Pe^{\imath\phi_\text{mod}(\vec{\boldsymbol{\alpha}})}\big)\star \hat{m}\Big)\Big(\big(\mathbb{I}_Pe^{-\imath\phi_\text{mod}(\vec{\boldsymbol{\alpha}})}\textcolor{black}{\phi}\big)\star \bar{\hat{m}}\Big)\bigg]\label{f2}
\end{equation}
where $\bar{.}$ means the complex conjugate. The linear intensity is a 2D map corresponding to the perfectly linear response of the WFS regarding to the phase $\phi$. By using the tip/tilt nature of the modulation \eqref{ttm}, it becomes possible to reveal the spatial variations of $I_\text{linear}$:
\begin{equation}
I_\text{linear}(\phi)|_{\vec{\textbf{R}}}=2 \textbf{Im}\bigg[\int_{\mathbb{R}^4} \text{d}^2\vec{\textbf{r'}} \text{d}^2\vec{\textbf{r}}~~\mathbb{I}_P|_{\vec{\textbf{r'}}} (\mathbb{I}_P\phi)|_{\vec{\textbf{r}}}\hat{m}|_{\vec{\textbf{R}}-\vec{\textbf{r'}}}\bar{\hat{m}}|_{\vec{\textbf{R}}-\vec{\textbf{r}}}  \int_{\mathbb{R}^2} \text{d}^2 \vec{\boldsymbol{\alpha}}w|_{\vec{\boldsymbol{\alpha}}}e^{\frac{2\imath\pi}{\lambda}\vec{\boldsymbol{\alpha}}.(\vec{\textbf{r'}}-\vec{\textbf{r}})}\bigg]\label{trollo}
\end{equation}
where $\vec{\textbf{R}}$ is the position vector in the intensity plane and $\vec{\textbf{r}}$ and $\vec{\textbf{r'}}$ the two integration variables coming from convolution products. We realize that, by integrating along tip/tilt amplitudes $\vec{\boldsymbol{\alpha}}$, the 2D Fourier transform of the weighting function appears:
\begin{equation}
I_\text{linear}(\phi)|_{\vec{\textbf{R}}}=2 \textbf{Im}\bigg[\int_{\mathbb{R}^4} \text{d}^2\vec{\textbf{r'}} \text{d}^2\vec{\textbf{r}}~~\mathbb{I}_P|_{\vec{\textbf{r'}}} (\mathbb{I}_P\phi)|_{\vec{\textbf{r}}}\\\hat{m}|_{\vec{\textbf{R}}-\vec{\textbf{r'}}}~\bar{\hat{m}}|_{\vec{\textbf{R}}-\vec{\textbf{r}}} \hat{w}|_{\vec{\textbf{r'}}-\vec{\textbf{r}}}\bigg]
\label{troll}
\end{equation}
Such a result has already been observed for the 1D Pyramid WFS with linear tip/tilt modulation in \cite{verinaud2004}. Its 2D generalization for any kind of modulations and masks is here allowed thanks to the fact that weighting function is considered as depending on tip/tilt amplitude vector $\vec{\boldsymbol{\alpha}}$ and not on the time variable. 

\subsection{Fourier based WFS's kernel}

At this point, \eqref{troll} is sufficiently developed to be interpreted as an integral transform performed on the input phase. Indeed, we can write
\begin{equation}
I_\text{linear} (\phi)|_{\vec{\textbf{R}}} =\int_{\mathbb{R}^2} \text{d}^2\vec{\textbf{r}} ~ \textbf{K}|_{\vec{\textbf{R}};\vec{\textbf{r}}}~ (\mathbb{I}_P\phi)|_{\vec{\textbf{r}}} \label{fondaa}
\end{equation}
where the \emph{Kernel} $\textbf{K}$ equals to:
\begin{equation}
\textbf{K}|_{\vec{\textbf{R}};\vec{\textbf{r}}} = 2\textbf{Im}\left[\bar{\hat{m}}|_{\vec{\textbf{R}}-\vec{\textbf{r}}}\int_{\mathbb{R}^2}\text{d}^2\vec{\textbf{r'}}~
\mathbb{I}_P|_{\vec{\textbf{r'}}}~\hat{m}|_{\vec{\textbf{R}}-\vec{\textbf{r'}}}~\hat{w}|_{\vec{\textbf{r'}}-\vec{\textbf{r}}}\right]\label{KKK}
\end{equation}
We observe that such a kernel depends on the WFS optical characteristics.  More precisely, we see that it actually depends on "pupil plane" functions: $\hat{m}$, $\hat{w}$ and  $\mathbb{I}_P$. Moreover, it is worth noticing that this kernel may be also understood as the continuous interaction matrix with respect to the natural basis of the direct phase space, i.e. the "Dirac phase basis". Such a fact is obvious when looking at  \eqref{fondaa}. 
%\begin{equation}
%\mathcal{B}_\delta= \left\{\phi_{x,y}^\delta : (a,b) \rightarrow  \delta(a-x) \delta(b-y)~;~ (x,y)\in \mathbb{R}^2 \right\}\label{DIRAC}
%\end{equation}
%The Kernel \textbf{K} is:
%\begin{equation}
%\textbf{K}|_{X,Y;x,y}=I_\text{linear} (\phi^\delta_{x,y})|_{X,Y}\label{Kd}
%\end{equation}
This property of the Kernel is fundamental since the decomposition of an arbitrary phase on the Dirac phase basis is absolutely trivial. In other word, $\textbf{K}$ is a very general descriptor of a WFS since it allows to compute naturally the WFS's response to any set of phases.

Finally, since most of the Fourier-based WFSs work without modulation, we give the Kernel when the tip/tilt modulation mirror generates a motionless and centered focal spot. In this case, the weighting function is a Dirac function. Subsequently, its Fourier transform equals to the identity function. The resulting kernel is: 
%\begin{equation}
%w=\delta~~~~~\hat{w}=\mathbb{I}
%\end{equation}
%It implies that:
\begin{equation}
\text{Without modulation:}~~~~~\textbf{K}|_{\vec{\textbf{R}};\vec{\textbf{r}}} = 2\textbf{Im}\Big[\bar{\hat{m}}|_{\vec{\textbf{R}}-\vec{\textbf{r}}}~(\mathbb{I}_P \star \hat{m})|_{\vec{\textbf{R}}}\Big]
\end{equation}
In that specific case, we observe that the matrix associated to this kernel may be understood as an Hadamard's product between a vertical matrix and a circulant one (see Appendix \ref{tata}). 

As a conclusion we will remember that the Kernel of a Fourier based Wave Front Sensor may be written as a real function of three "pupil plane" quantities and two spatial vectors, one in the input space the other in the output space. 
\begin{equation}
\textbf{K}(\hat{m},\hat{w},\mathbb{I}_P)|_{\vec{\textbf{R}};\vec{\textbf{r}}}
\end{equation}
In other words, for a given optical configuration, \textbf{K} is a continuous 2D matrix. Such a fact explains why the Kernel requires in general, onerous simulations in terms of computational time. Using it to lead an exhaustive study of Fourier based WFSs seems thus unrealistic. Moreover, despite the elegance of its expression, the input/output relation \eqref{fondaa} remains difficult to invert directly; \textbf{K} has to have additional properties to be exploited effectively. That is the goal of the next part where we will work with \emph{convolutional} Kernels, i.e. Kernels which are not function of $\vec{\textbf{R}}$ \emph{and} $\vec{\textbf{r}}$ but only of their difference $\vec{\textbf{R}}-\vec{\textbf{r}}$.

\section{Convolutional model} \label{convy}

From now, we will study the consequences of having a Kernel which would be convolutional. In that case, it exists a function, that we call $\textbf{IR}$ for \textbf{I}mpulse \textbf{R}esponse which allows to write:
\begin{equation}
\textbf{K}|_{\vec{\textbf{R}};\vec{\textbf{r}}}=\textbf{IR}|_{\vec{\textbf{R}}-\vec{\textbf{r}}}\label{KeK}
\end{equation}
The input/output relation of the WFS becomes convolutional:
\begin{equation}
I_\text{linear}(\phi) = (\mathbb{I}_P\phi) \star \textbf{IR}\label{convovov} 
\end{equation}
It is worth noticing that convolutional systems are also called shift invariant. Indeed, we observe that if $\mathcal{T}$ is a translation operator, the convolution implies that $I_\text{linear}(\mathcal{T}[\phi])=\mathcal{T}[I_\text{linear}(\phi)]$. In other words, knowing the WFS's response for a given phase is enough to know the response to any translation of this phase. Such a result explain why a convolutional Kernel may be more rapidly computed: the associated matrix is pure circulant, i.e. a strongly redundant matrix. On the other hand,  \eqref{convovov} clarifies the notation "Impulse Response", indeed this quantity corresponds to the WFS response when the phase is a pure impulse, i.e. a centered Dirac.

\subsection{Required assumptions}

\begin{figure}
    \centering
    \includegraphics[width=4cm]{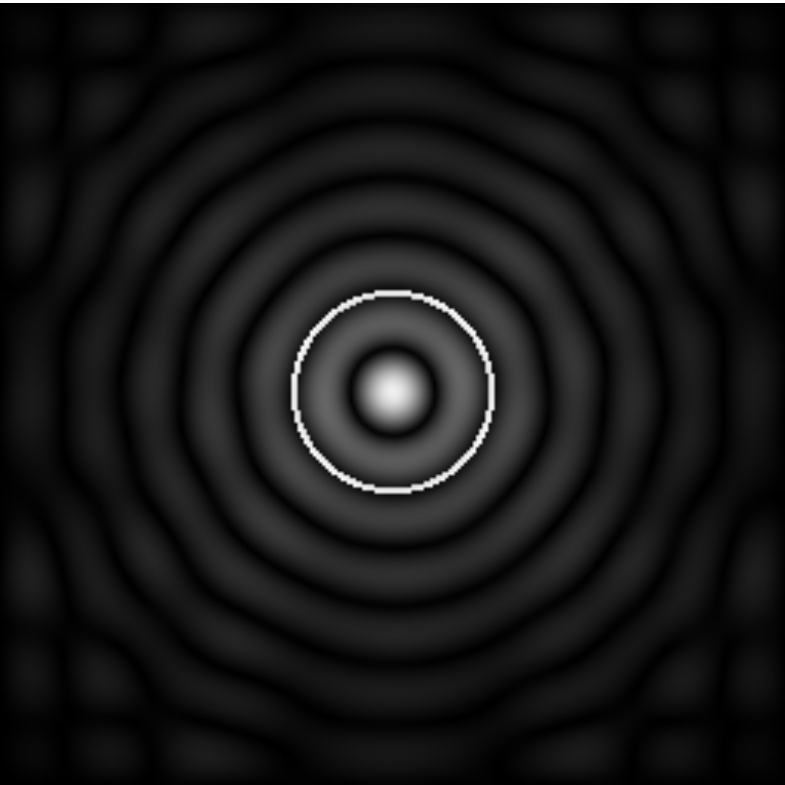}
    
    $\hat{w}$ and $\partial\mathbb{I}_P$
    \caption{Overlay of the Fourier transform of the weighting function (background) and the pupil edge (white circle). We consider here a circular pupil and a ring modulation with a modulation radius ensuring \eqref{import}.  \label{wdi}}
    
\end{figure}

Since Kernel is not in general convolutional, we are now interested on the assumptions needed to have such a Kernel. 
Our starting equation is  \eqref{KKK}. 
By taking a closer look at it, we observe that the main difficulty comes from the combined presence of functions depending on $\vec{\textbf{r'}}$ and $\vec{\textbf{r'}}-\vec{\textbf{r}}$ variables; this fact is an obvious obstacle to shift invariance. In other words, converging toward the convolutional model consists in eliminating one of two previous dependency. To do so, we make a Taylor's development of the pupil indicator function $\mathbb{I}_P|_{\vec{\textbf{r'}}}$ around $\vec{\textbf{r'}}-\vec{\textbf{r}}$:
\begin{equation}
\mathbb{I}_P|_{\vec{\textbf{r'}}} = \mathbb{I}_P|_{\vec{\textbf{r'}}-\vec{\textbf{r}}} + \vec{\textbf{r}}\cdot \left(\vec{\nabla} \mathbb{I}_P \big|_{\vec{\textbf{r'}}-\vec{\textbf{r}}}\right) +...
\end{equation}
where $\vec{\nabla}$ is the 2D gradient operator. Such a development allows to write the Kernel as:
\begin{equation}
\textbf{K}|_{\vec{\textbf{R}};\vec{\textbf{r}}} = 2\textbf{Im}\big[\bar{\hat{m}}(\hat{m}\star \big(\hat{w}\mathbb{I}_P)\big)\big]\big|_{\vec{\textbf{R}}-\vec{\textbf{r}}} 
\\ + 2 \vec{\textbf{r}} \cdot \textbf{Im}\big[\bar{\hat{m}}(\hat{m}\star \big(\hat{w}\vec{\nabla} \mathbb{I}_P)\big)\big]\big|_{\vec{\textbf{R}}-\vec{\textbf{r}}} +...\label{rererere}
\end{equation}
The convolutional model consists in keeping the first term (which has a convolutional form) and neglecting the next ones (which do not).
In most cases, such a simplification is not exact and constitute an approximation of the linear model. We call this assumption the \emph{sliding pupil approximation} since it corresponds to the following assumption:
\begin{equation}
\mathbb{I}_P|_{\vec{\textbf{r'}}} \approx \mathbb{I}_P|_{\vec{\textbf{r'}}-\vec{\textbf{r}}}\label{pupgliss}
\end{equation}
Nevertheless, we note that   \eqref{pupgliss} is not an assumption in two cases. First one is a bit unrealistic but worthy of mention: it is the infinite pupil case. Indeed we observe that $\mathbb{I}_P = \mathbb{I}$ implies a pure convolutional kernel. The second case is much more interesting since it shows how it is possible to choose the tip/tilt modulation in order to improve the accuracy of the convolutional model.  It consists in using a weighting function $w$ which ensures:
\begin{equation}
\hat{w} \partial \mathbb{I}_P = 0 \label{import}
\end{equation}
where $\partial \mathbb{I}_P$ corresponds to the area where the entrance pupil has discontinuities. In other words, a Fourier-based WFS is a convolutional sensor if the Fourier Transform of the weighting function $\hat{w}$ is barely null on the edge of the pupil. Such a condition is actually quite simple to reach. For a circular pupil and a ring modulation, for instance, $\hat{w}$ is the first Bessel function of the first kind: $J_0$. In order to ensure \eqref{import}, we need adapt the modulation radius in such a way that the pupil radius corresponds to a Bessel function's zero. An example is shown on Fig. \ref{wdi}. In other words, it would be possible to improve the accuracy of the convolutional model on the condition to properly set the modulation radius.  

\subsection{Impulse Response and Transfer function}

\begin{figure}[htbp]
\centering
\includegraphics[width=12cm]{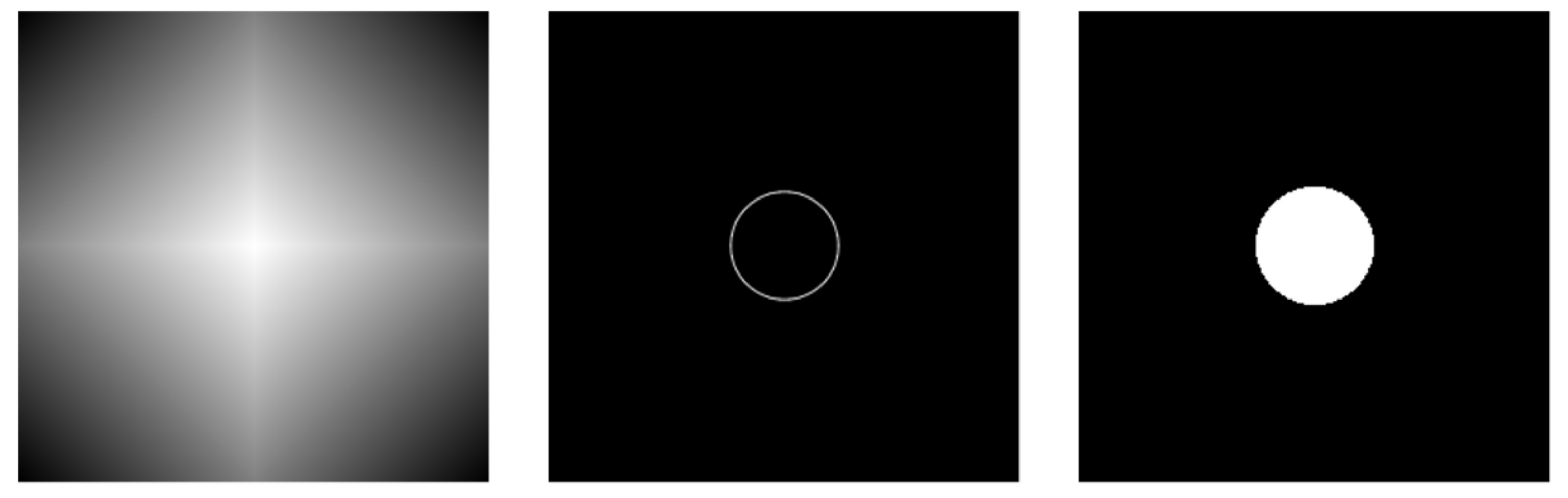}

arg($m$) ~~~~~~~~~~~~~~~~~~~~~~~~~~~~~~~ $w$ ~~~~~~~~~~~~~~~~~~~~~~~~~~~~~~~~~ ~~~~$\mathbb{I}_P$
\caption{Optical parameters needed to compute the Impulse Response. From left to right: argument of the transparency function, weighting function and pupil's indicator function. We chose the particular case of the 4-sided Pyramid with a ring modulation and a circular entrance pupil.\label{tuy}}
\end{figure}

Assuming the sliding pupil approximation, it is possible to identify the WFS's Impulse Response \textbf{IR} in  \eqref{rererere}:
\begin{equation}
\textbf{IR} =  2\textbf{Im}\big[\bar{\hat{m}}\big(\hat{m} \star (\hat{w}\mathbb{I}_P)\big)\big]\label{IRIRIR}
\end{equation} 
The Impulse Response is a concise way to characterize a Fourier-based WFS while assuming it is shift invariant. We observe that the computational cost which is required to calculate the \textbf{IR} depending on the optical parameters (see Fig. \ref{tuy})  is absolutely negligible compared to calibration matrices' approaches. Finally, we emphasize on the fact that the circulant matrix $\textbf{IR}|_{\vec{\textbf{R}}-\vec{\textbf{r}}}$ may be seen as the response of the sensor with respect to the Dirac phase basis, i.e. the natural basis of the phase space. 
\begin{figure}[htbp]
\centering
\includegraphics[trim = 1.5cm 1.5cm 1.5cm 1.5cm, clip,width=4cm]{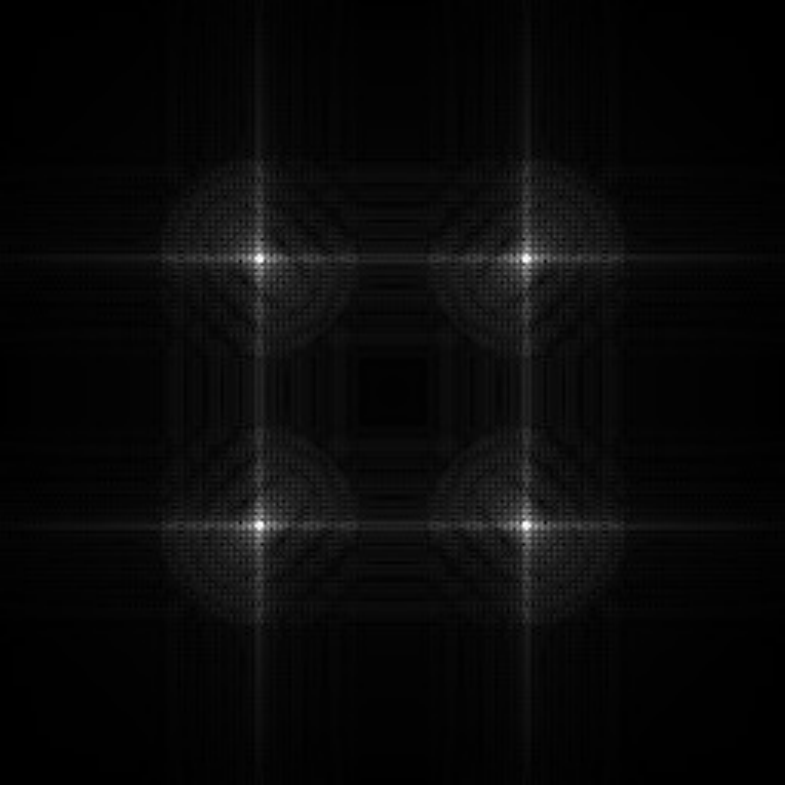}

\textbf{IR}

\caption{Impulse response corresponding to the optical parameters of Fig. \ref{tuy}. \label{ert}}
\end{figure}
To get the response of the WFS in the phase spatial frequencies space, i.e. in a focal plane, we just have to consider the 2D Fourier transform of the impulse response that we call Transfer Function \textbf{TF}:
\begin{eqnarray}
\textbf{TF} \equiv \widehat{\textbf{IR}}
\end{eqnarray}
For the sake of clarity, we do not give here the full expression of this important quantity. The reader is referred to the paragraph \ref{JUJU} for a comprehensive formulation of the \textbf{TF} depending on the optical parameters. Indeed, we will firstly generalize the convolutional model to more realistic wave front sensing contexts.

\section{Wave front sensing in presence of residual phases}

In this part, we complicate the sensing context by studying a typical Adaptive Optics case : the presence in the pupil plane of residual phases which are distinct from the phase-to-be-measured. They may be static (as for instance the vast majority of Non Common Path Aberrations (NCPA)) or dynamic (as for instance uncorrected turbulent phase). In the first case, we call these terms $\phi_{\text{s}}$ for \textbf{s}tatic residuals. In the same way, we call the \textbf{d}ynamic ones $\phi_{\text{d}}$ and we assume their statistical behavior to be known. In particular, the notation $\big\langle.\big\rangle $ means "average on the residual phases statistics".

The residual phases are taken into account exactly as the tip/tilt modulation, i.e. by adding phases to the incoming field. We thus replicate the calculations of paragraph \ref{fffr} while assuming the sliding pupil approximation \eqref{pupgliss} and using the linearity of the average operation. The WFS's kernel associated to this new situation is:
\begin{equation}
\textbf{K}|_{\vec{\textbf{R}};\vec{\textbf{r}}} = 2\textbf{Im}\left[\bar{\hat{m}}|_{\vec{\textbf{R}}-\vec{\textbf{r}}}\int_{\mathbb{R}^2}\text{d}^2\vec{\textbf{r'}}~
\hat{m}|_{\vec{\textbf{R}}-\vec{\textbf{r'}}}~(\hat{w}\mathbb{I}_P)|_{\vec{\textbf{r'}}-\vec{\textbf{r}}}~\textbf{R}_\text{s}|_{\vec{\textbf{r'}};\vec{\textbf{r}}}\textbf{R}_\text{d}|_{\vec{\textbf{r'}};\vec{\textbf{r}}}\right] \label{tytyyt}
\end{equation}
where functions $\textbf{R}_s$ and $\textbf{R}_d$ characterize the  \textbf{s}tatic and \textbf{d}ynamic \textbf{R}esidual phases:
\begin{eqnarray}
\textbf{R}_{\text{s}}|_{\vec{\textbf{r'}};\vec{\textbf{r}}} &\equiv&   e^{{-\imath\phi_{\text{s}}}|_{\vec{\textbf{r}}}}~e^{{\imath\phi_{\text{s}}}|_{\vec{\textbf{r'}}}} \label{RS} \\
\textbf{R}_{\text{d}}|_{\vec{\textbf{r'}};\vec{\textbf{r}}} &\equiv&  \big\langle  e^{{-\imath\phi_{\text{d}}}|_{\vec{\textbf{r}}}}~e^{{\imath\phi_{\text{d}}}|_{\vec{\textbf{r'}}}} \big\rangle 
\end{eqnarray}
We observe unfortunately that residual phases break the convolutional model: it is impossible to write  \eqref{tytyyt} as a convolutional relation without doing supplementary assumptions. 

\subsection{Dynamic residuals}

We firstly consider dynamic residuals. In an Adaptive Optics context, we can assume that they are Gaussian-distributed and zero-mean phase, it allows us to simplify the function $\textbf{R}_\text{d}$ into:
\begin{equation}
\textbf{R}_\text{d}|_{\vec{\textbf{r'}};\vec{\textbf{r}}}= e^{-\frac{1}{2}\big\langle (\phi_{\text{d}}|_{\vec{\textbf{r'}}}-\phi_{\text{d}}|_{\vec{\textbf{r}}})^2\big\rangle}= e^{-\frac{1}{2}\textbf{D}|_{\vec{\textbf{r'}};\vec{\textbf{r}}}}
\end{equation}
where $\textbf{D}$ is the structure function of the residual phases \cite{tatar1961}. This quantity is much more easy to compute and well-known concerning the full or residual atmospheric turbulence. 
In order to go even further, we can use the eventual stationarity of the dynamic residuals which says that the structure function only depends on the distance between the points where it is computed:
\begin{equation}
\textbf{D}|_{\vec{\textbf{r'}};\vec{\textbf{r}}}=\textbf{D}|_{\vec{\textbf{r'}}-\vec{\textbf{r}}} \implies \textbf{R}_\text{d}|_{\vec{\textbf{r'}};\vec{\textbf{r}}} = \textbf{R}_\text{d}|_{\vec{\textbf{r'}}-\vec{\textbf{r}}}
\end{equation}
Thus, the set of assumptions "Gaussian-distributed + zero-mean + stationarity" allows to make $\textbf{R}_d$ function of $\vec{\textbf{r'}}-\vec{\textbf{r}}$ only. In other words, if these assumptions are valid, sensing in presence of dynamic residuals is compatible with the convolutional kernel. It also means that obtaining \textbf{K} does not require doing a large number of statistical realizations : knowing the structure function is enough. Such a fact implies a significant acceleration of numerical simulations. Finally, it is worth noticing that to take dynamic residuals into account, we just have to change the weighting function by using the structure function \textbf{D}:
\begin{equation}
\hat{w}\mathbb{I}_P~~\rightarrow~~\hat{w}\mathbb{I}_Pe^{-\frac{1}{2}\textbf{D}} \label{Dphh}
\end{equation}
Such a fact is, afterwards, quite logical. The tip/tilt modulation itself may be understood as a "dynamic residual phases"; its weighting function is a way to characterize its statistics. 

\subsection{Static residual}

We are now interested in static residual. We observe in  \eqref{RS} that $\textbf{R}_s$ does not depend on $\vec{\textbf{r'}}-\vec{\textbf{r}}$ only. Thus the kernel is not convolutional due to the presence of static residual. To be able to use the convolutional model, we perform a Taylor's development of the  phases difference in  \eqref{RS} around $\vec{\textbf{r'}}-\vec{\textbf{r}}$:
\begin{eqnarray}
\phi_{\textsc{s}}|_{\vec{\textbf{r'}}} - \phi_{\textsc{s}}|_{\vec{\textbf{r}}}  &=&  (\vec{\textbf{r'}}-\vec{\textbf{r}})\cdot\left(\vec{\nabla} \phi_{\textsc{s}} \big|_{\vec{\textbf{r'}-\vec{\textbf{r}}}}\right) 
+ ...|_{\vec{\textbf{r'}};\vec{\textbf{r}}}\\
&=& \mathcal{D}\phi_s|_{\vec{\textbf{r'}}-\vec{\textbf{r}}}+...|_{\vec{\textbf{r'}};\vec{\textbf{r}}}\label{Euler}
\end{eqnarray}
where $\mathcal{D}$ is a \textbf{d}ifferential operator defined as: 
\begin{equation}
\mathcal{D}\phi|_{\vec{\textbf{r}}}\equiv  \vec{\textbf{r}}\cdot\vec{\nabla}\phi|_{\vec{\textbf{r}}}
\end{equation}
We observe that keeping the first term in  \eqref{Euler} makes $\textbf{R}_s$ a function of $\vec{\textbf{r'}}-\vec{\textbf{r}}$ only. Generally this approximation is not valid but it is worth noticing that the only static residual phases for which it is not an approximation are pure tip/tilt aberrations. Physically, it means that the only static residuals which are rigorously compatible with the convolutional model are tip/tilt.  This is not that surprising: adding such a static phase in the pupil plane is absolutely equivalent to using a shifted weighting modulation. Nevertheless, it can be shown that the first term of  \eqref{Euler} are predominant compared to the next ones as soon as $\phi_s$ only contains low order aberrations.
To summarize, if this "low order approximation" is valid the convolutional model becomes usable again; taking into account static residual just consists in adapting the weighting function: 
\begin{equation}
\hat{w}\mathbb{I}_Pe^{-\frac{1}{2}\textbf{D}}~~\rightarrow~~\hat{w}\mathbb{I}_Pe^{-\frac{1}{2}\textbf{D}}e^{\imath\mathcal{D}\phi_s} %\label{Dphh}
\end{equation}

\subsection{Effective Modulation}\label{JUJU}

\begin{figure}[htbp]
\centering
\includegraphics[scale=0.27]{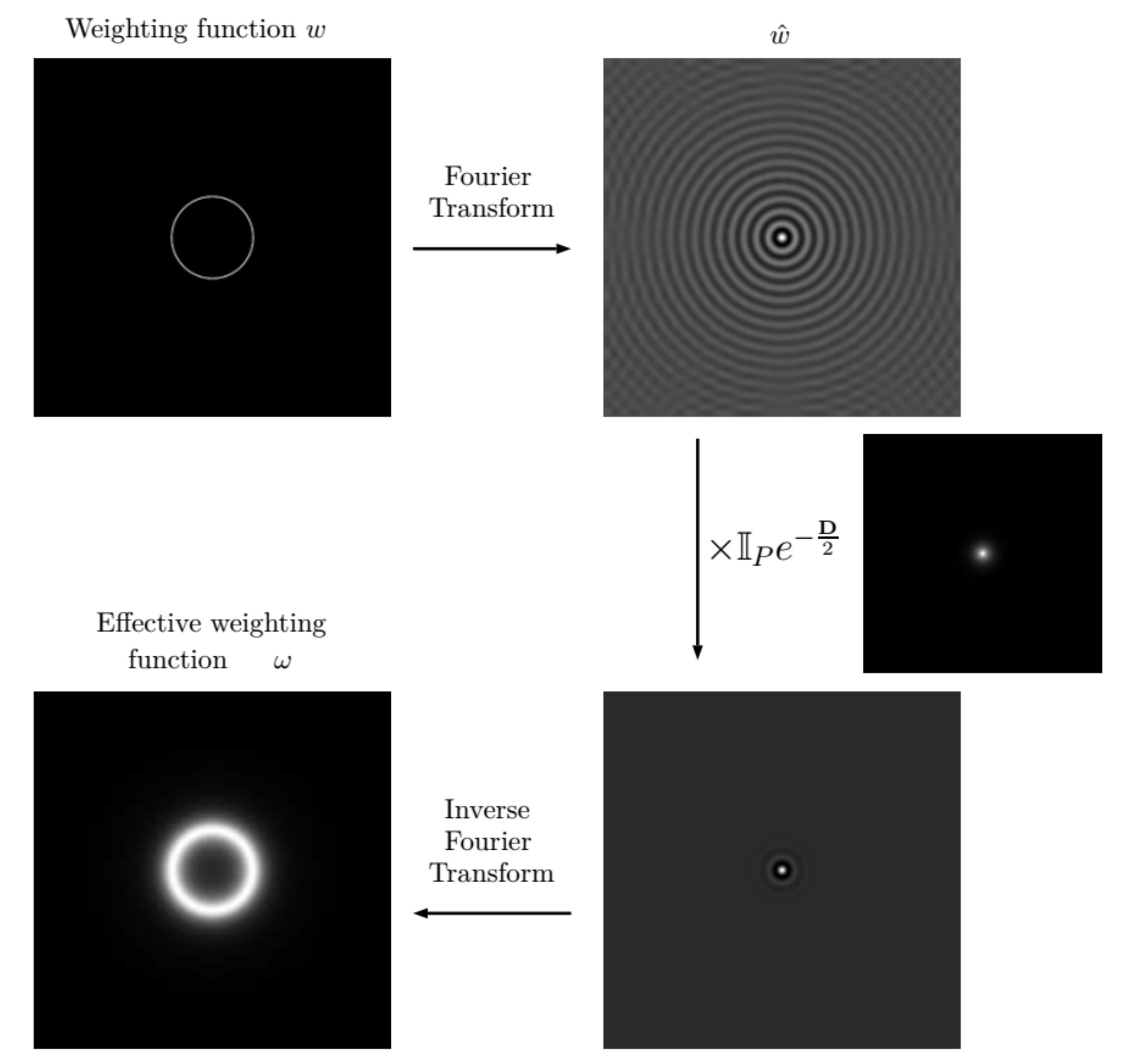}

\caption{Numerical method to compute the effective weighting function in presence of dynamic residual phases. We consider in this example a ring modulation with a modulation radius of 7 $\lambda/D$ and a typical Kolmogorov turbulence structure function with $r_0=D/4$. \label{weff}}
\end{figure}

We are now ready to give the Impulse Response in presence of dynamic and static residual phases. To do so, we examine the convolutional Kernel when the three following assumptions are valid: 
\begin{itemize}
\item "Sliding pupil approximation" about the pupil geometry. 
\item "Gaussian-distribution + zero-mean + stationarity" about the dynamic residuals statistics. 
\item "Low order static aberration approximation" about static residual phase. 
\end{itemize}
\begin{equation}
\textbf{K}|_{\vec{\textbf{R}};\vec{\textbf{r}}} = 2\textbf{Im}\left[\bar{\hat{m}}|_{\vec{\textbf{R}}-\vec{\textbf{r}}}\int_{\mathbb{R}^2}\text{d}^2\vec{\textbf{r'}}~
\hat{m}|_{\vec{\textbf{R}}-\vec{\textbf{r'}}}~\left(\hat{w}\mathbb{I}_Pe^{-\frac{1}{2}\textbf{D}}e^{\imath\mathcal{D}\phi_s}\right)|_{\vec{\textbf{r'}}-\vec{\textbf{r}}}\right]
\end{equation}
At first sight, this Kernel depends on 5 distinct quantities: 3 related to the parameters of the optical Fourier filtering system ($m$,$w$,$\mathbb{I}_P$) and 2 which characterizes the residual phases ($\textbf{D}, \mathcal{D}\phi_s$) but it is actually possible to reduce drastically this complexity by introducing a concise quantity that we call $\omega$ and which is defined via its Fourier transform:
\begin{equation}
\hat{\omega} \equiv \hat{w}\mathbb{I}_Pe^{-\frac{1}{2}\textbf{D}}e^{\imath \mathcal{D}\phi_\text{s}}
\end{equation}
Physically, $\omega$ may be seen as an \emph{effective modulation weighting function} when taking into account the finite size of the pupil and the residual phases.  To be convinced of this, we will lead a small mental experiment by including just before the filtering mask, an imaginary camera. This one will allow us to visualize in the focal plane the effects of the tip/tilt modulation and residual phases. 
We start with the most basic sensing context, i.e. an infinite pupil without any residual phases.  In that case, the fictive camera will record the weighting function $w$, exactly as if $w$ were the shape of an extended object. Thus the effective weighting function rigorously equals to the tip/tilt weighting function, cf.  \eqref{1}. Unfortunately, this case is not realistic: the pupil cannot have an infinite size, this one necessarily windows the incoming field which will result in a blurring of the tip/tilt weighting function. More precisely, $w$ is convolved with the Point Spread Function associated to the pupil. In the Fourier space, it exactly corresponds to  \eqref{2}. If we add now a static residual $\phi_s$, the Point Spread Function will be degraded. Thus, a new term has to be added in $\omega$; it is the  \eqref{3}. We can repeat this argumentation for dynamic residual phases only  \eqref{4}. The most general case corresponds to  \eqref{5}; it allows to understand the sensing in presence of dynamic and static residual phases as a tip/tilt modulation using the effective weighting function $\omega$. An example is given on Fig. \ref{weff} which shows how to compute the effective weighting function when sensing in open loop on a typical turbulent atmosphere. We may observe a blurring of the original weighting function due to the low-pass filtering induced by the $\mathbb{I}_Pe^{-\frac{1}{2}\textbf{D}}$ term. Physically, the residual phases act like a supplementary modulation.
\begin{eqnarray}
&\text{Infinite Pupil} & \hat{\omega} = \hat{w}\label{1}\\
&\text{Finite Pupil} &  \hat{\omega} = \hat{w}\mathbb{I}_P\label{2}\\
&\text{Finite Pupil + static residual} & \hat{\omega} = \hat{w}\mathbb{I}_Pe^{\imath \mathcal{D}\phi_\text{s}}\label{3}\\
&\text{Finite Pupil + dynamic residuals} & \hat{\omega} = \hat{w}\mathbb{I}_Pe^{-\frac{\textbf{D}}{2}}\label{4}\\
&\text{Finite Pupil + dynamic \& static } & \hat{\omega} = \hat{w}\mathbb{I}_Pe^{\imath \mathcal{D}\phi_\text{s}}e^{-\frac{\textbf{D}}{2}}\label{5}
\end{eqnarray}
In conclusion, to characterize a WFS in the convolutional model, a generic Impulse Response depending on the mask $m$ and the effective weighting function $w$ only is sufficient: 
\begin{equation}
\textbf{IR} =  2\textbf{Im}\left[\bar{\hat{m}}(\hat{m\omega})\right]
\end{equation}
To take into account the tip/tilt modulation, the pupil geometry and the residual phases, we just have to use  \eqref{2} to \eqref{5}. Such a result is also valid for the Transfer Function which has a concise expression when depending on $m$ and $w$.
\begin{equation}
\textbf{TF}=\imath(\hat{\hat{m}}\star\overline{m\omega}-\overline{m}\star\hat{\hat{m\omega}})\label{symsym}
\end{equation}
Indeed, we observe that the \textbf{TF} only requires basic mathematical operations: conjugation, convolution and double Fourier transform (i.e. symmetry operation). It is worth noticing that \eqref{symsym} is even more clear if the optical system is centro-symmetric. (It is for instance the case of the 4-faces Pyramid WFS with a ring modulation and isotropic residual phases.) As a matter of fact, we get:
\begin{equation}
\textbf{TF}=2\textbf{Im}\big[m\star\overline{m\omega}\big] \label{essai}
\end{equation}
The computational simplicity of  \eqref{symsym} or  \eqref{essai} is especially interesting because of its link with the WFS's sensitivity regarding to the phase spatial frequencies (for more details, see appendix \ref{ujuj}). Indeed this quantity is the essential to know how noises are propagating in AO loops, see for instance \cite{carlos2018} or to build efficient convolution based phase reconstructors \cite{shatok2017}.

\section{Conclusion}

This paper is dedicated to a natural extension of the theoretical work done in \cite{Fauvarque16} about Fourier-based WFSs.  Using the mathematical framework describing these sensors, we showed that they may be understood as 2D integral operators which are completely characterized by their Kernel.  

In the first part, we provided an explicit expression of this quantity depending on the WFS's optical parameters: pupil geometry, weighting function of the tip/tilt modulation and filtering mask and showed that it can be understood as the continuous calibration matrix with respect to the Dirac phase basis. Nevertheless, in spite of the generality of the kernel's expression, this one remains difficult to calculate numerically because of the computational complexity introduced by the tip/tilt modulation. 

The particular case of convolutional kernels, which correspond to highly redundant circulant matrices, solves this problem. In this case, the WFS's output equals the convolution product between the phase-to-be-measured and a quantity called the Impulse Response of the sensor. It was found that this quantity directly depends on the optical parameters and then allows to calculate performance criterion like the sensor's sensitivity with respect to the spatial frequencies in a much faster way than existing numerical simulations. 
The convolutional model then represents an ideal tool to compare and optimize in an exhaustive way the Fourier based WFSs, as for instance the very abundant Pyramid WFSs class. Indeed, if such a study has already been done for the filtering masks ($n$-faces Pyramids \cite{fauvarque2017JATIS}, Axicon \cite{Akondi2013a, Akondi2014, Clare2003, Clare2005}; Flattened Pyramid \cite{Fauvarque2015}; masks with manufacturing errors, etc.) the tip/tilt modulation still remained to be explored due to the computing time it required.  Thanks to the convolutional model, all the 1D modulations (ring, square, etc.) but also 2D modulations (disc, Gaussian, etc.) are easily taken into account. This also opens the way to random modulation or sensing with extended objects as Laser Guide Stars.

In Appendix B, we showed how the convolutional model could be applied to the classical 4-faces pyramid when using the slopes maps rather than differential intensity as output of the WFS. We also proved that the sensitivity derived from the convolutional model is in total agreement with experimental measurements \cite{bond2018}.

Since the convolutional model is an approximation, we paid special attention to the required assumptions. In particular, we showed that the infinite pupil approximation \cite{poyneer2003, quiross2010, shatok2017} is not the only way to justify the use of that model: a weaker assumption called the sliding pupil assumption allows to be closer to the real model since it does take into account the pupil geometry's. These results suggests that advances may be done in deconvolution-based phase reconstruction by adapting algorithms to sliding pupil approximation.

Finally, we extended the convolutional model to more realistic sensing contexts. The goal was to know how to adapt the convolutional Kernel in presence of residual phases which differ from the phase-to-be-measured. Two cases were addressed: static residual to handle for instance Non Common Path Aberrations and dynamic residuals to take into account non-corrected turbulent phases. We showed that under realistic assumptions about residuals statistics, one only has to modify the modulation weighting function, as for instance by introducing the structure function of the residual phases. These results are an important step in the theoretical understanding of wave front sensing in presence of residuals. Indeed, they demonstrate that these "non pure tip/tilt phases" which vary in the pupil plane during the sensing may still be described via the tip/tilt modulation weighting function. Consequently, the convolutional model remains valid when sensing with residuals. 

\medskip

%By way of conclusion, we would like to refer the reader to two forthcoming papers which use convolutional model to tackle two hot topics concerning the Adaptive Optics of the future extremely large telescopes. 

%The first one \cite{carlos2018} takes advantage of the simplicity of the convolutional input/output relation to provide a comprehensive analysis of the performance limits of AO/high-contrast imagers which uses Pyramid WFS.  

This result opens the path to detailed studies of the so-called "optical gain". These "optical gains" correspond to variations of the Pyramid WFS sensitivity in presence of partially-corrected turbulent phases. A solution to this problem consists in adjusting modal gains in the AO loop to compensate the loss of sensitivity \cite{Kork08}. The determination of the modal gains can be derived from empirical methods \cite{deo2018}, and the convolutional model now provides an alternative approach to estimate the optimal optical gain directly.

\section*{Funding} This project has been supported by the LABEX FOCUS, the WOLF ANR project (grant No 183707), the Action Sp\'ecifique Haute R\'esolution Angulaire (ASHRA) of CNRS/INSU co-funded by CNES and the European Union's Horizon 2020 dearch and innovation programme under grant agreement (grant No 730890). It is also partly funded by the French Aerospace Lab (ONERA) in the frame of the VASCO Research Project and by the Laboratoire d'Astrophysique de Marseille (LAM). C. Correia received support from A*MIDEX (project no. ANR-11- IDEX-0001- 02) funded by the “Investissements d’Avenir" French Government program, managed by the French National Research Agency (ANR).

\section*{Acknowledgments} Victoria Hutterer (Johannes Kepler University, Linz) is acknowledged for her careful rereading and her judicious advice.

\appendix

\section{Sensitivity with respect to phase spatial frequencies}\label{ujuj}

This appendix aims to show the link between the Transfer Function and the sensitivity  with respect to phase spatial frequencies.  Such a basis describes phases in terms of Sines and Cosines:
\begin{equation}
 \Big\{ \phi_{\vec{\textbf{k}}}^{\cos}~:\vec{\textbf{r}}\rightarrow\cos\left(2\pi\vec{\textbf{k}}.\vec{\textbf{r}}\right) ~~ \text{and} ~~\phi_{\vec{\textbf{k}}}^{\sin}~:\vec{\textbf{r}}\rightarrow\sin\left(2\pi\vec{\textbf{k}}.\vec{\textbf{r}}\right)~~\text{with}~~ \vec{\textbf{k}} \in \mathbb{R}^2\Big\}
\end{equation}
where $\vec{\textbf{k}}$ may be seen as a position vector in the spatial frequencies space. Usually, the sensitivity is defined thanks to a scalar which quantifies the ratio between WFS's output and input. For instance, the sensitivity with respect to a phase mode $\phi$ is defined as: 
\begin{equation}
\frac{||I_\text{linear}(\mathbb{I}_P\phi) ||_2}{||\phi||_2}
\end{equation}
We showed in \cite{Fauvarque16} that such a definition was consistent with \cite{guyon2005limits} and \cite{Rigaut92} approaches.
In our case, a more convenient way to represent the sensitivity consists in grouping the sine and cosine sensitivities into a unique number associated to $\vec{\textbf{k}}$. 
\begin{equation}
s|_{\vec{\textbf{k}}} \equiv\sqrt{\left|\left|I_\text{linear}(\mathbb{I}_P\phi^{\cos}_{\vec{\textbf{k}}})\right|\right|_2^2 + \left|\left|I_\text{linear}(\mathbb{I}_P\phi^{\sin}_{\vec{\textbf{k}}})\right|\right|_2^2}\label{ttereter}
\end{equation}
where we purposely did not write a scalar coefficient allowing to normalize these phases on the pupil support with respect to RMS norm. From now, we will assume that the convolutional model is valid to simplify this expression. We have for instance for the cos term:
\begin{eqnarray}
\left|\left|I_\text{linear}(\mathbb{I}_P\phi^{\cos}_{\vec{\textbf{k}}})\right|\right|_2^2 &=& \left|\left|(\mathbb{I}_P\phi^{\cos}_{\vec{\textbf{k}}})\star \textbf{IR}\right|\right|_2^2\\
&=&\left|\left|\widehat{\mathbb{I}_P\phi^{\cos}_{\vec{\textbf{k}}}} \times \textbf{TF}\right|\right|_2^2\\
&=& \left|\left|\left( \widehat{\mathbb{I}_P}\star \widehat{\phi^{\cos}_{\vec{\textbf{k}}}}\right)\times \textbf{TF}\right|\right|_2^2
\end{eqnarray}
where we used the Plancherel theorem. Moreover, we know that
\begin{equation}
\widehat{\phi_{\vec{\textbf{k}}}^{\cos}}=\frac{\delta_{\vec{\textbf{k}}}+\delta_{-\vec{\textbf{k}}}}{2}~~~~\text{and}~~~~
\widehat{\phi_{\vec{\textbf{k}}}^{\sin}}=\frac{\delta_{\vec{\textbf{k}}}-\delta_{-\vec{\textbf{k}}}}{2\imath}\label{frr}
\end{equation}
As a consequence, 
\begin{equation}
\left|\left|\left( \widehat{\mathbb{I}_P}\star \widehat{\phi^{\cos}_{\vec{\textbf{k}}}}\right)\times \textbf{TF}\right|\right|_2^2 = \frac{1}{4} \int_{\mathbb{R}^2} \text{d}\vec{\textbf{r}}~\Big|\widehat{\mathbb{I}_P}|_{\vec{\textbf{r}}-\vec{\textbf{k}}} \textbf{TF}|_{\vec{\textbf{r}}} + \widehat{\mathbb{I}_P}|_{\vec{\textbf{r}}+\vec{\textbf{k}}} \textbf{TF}|_{\vec{\textbf{r}}} \Big|^2 
\end{equation}
\begin{equation}
\left|\left|\left( \widehat{\mathbb{I}_P}\star \widehat{\phi^{\sin}_{\vec{\textbf{k}}}}\right)\times \textbf{TF}\right|\right|_2^2 = \frac{1}{4} \int_{\mathbb{R}^2} \text{d}\vec{\textbf{r}}~\Big|\widehat{\mathbb{I}_P}|_{\vec{\textbf{r}}-\vec{\textbf{k}}} \textbf{TF}|_{\vec{\textbf{r}}} - \widehat{\mathbb{I}_P}|_{\vec{\textbf{r}}+\vec{\textbf{k}}} \textbf{TF}|_{\vec{\textbf{r}}} \Big|^2 
\end{equation}
which gives
\begin{multline}
\left|\left|\left( \widehat{\mathbb{I}_P}\star \widehat{\phi^{\cos}_{\vec{\textbf{k}}}}\right)\times \textbf{TF}\right|\right|_2^2 + \left|\left|\left( \widehat{\mathbb{I}_P}\star \widehat{\phi^{\sin}_{\vec{\textbf{k}}}}\right)\times \textbf{TF}\right|\right|_2^2  =\\ \frac{1}{2}\int_{\mathbb{R}^2} \text{d}\vec{\textbf{r}}~\big|\widehat{\mathbb{I}_P}\big|^2\Big|_{\vec{\textbf{r}}-\vec{\textbf{k}}} \big|\textbf{TF}\big|^2\Big|_{\vec{\textbf{r}}} +\frac{1}{2}\int_{\mathbb{R}^2} \text{d}\vec{\textbf{r}}~\big|\widehat{\mathbb{I}_P}\big|^2\Big|_{\vec{\textbf{r}}+\vec{\textbf{k}}} \big|\textbf{TF}\big|^2\Big|_{\vec{\textbf{r}}} 
\end{multline}
With $|\widehat{\mathbb{I}_P}|^2$, we identify the \textbf{P}oint \textbf{S}pread \textbf{F}unction the diffraction limited imaging system. Since the indicator pupil function is real, the \textbf{PSF} is necessarily centro-symmetric. The sensitivity with respect to spatial frequencies finally equals to:
\begin{equation}
s|_{\vec{\textbf{k}}} = \sqrt{|\textbf{TF}|^2\star \textbf{PSF}}\Big|_{\vec{\textbf{k}}}\label{utje}
\end{equation}
This formula explains why the convolutional model is so efficient to explore the abundant world of Fourier-based WFSs. Indeed, it proves that it is possible to get very quickly an idea of the WFS's the response with respect to spatial frequencies; there is no need to build an interaction matrix via time demanding end-to-end codes, using our knowledge of the system's optical parameters ($m$, $w$ and $\mathbb{I}_P$) and the residuals nature ($\textbf{D}$ and $\mathcal{D}\phi_s$) in \eqref{symsym} and \eqref{utje} is enough.

\section{Application to the Pyramid WFS}\label{PyrPyr}

\begin{figure}[htbp]
\centering
\includegraphics[scale=0.8]{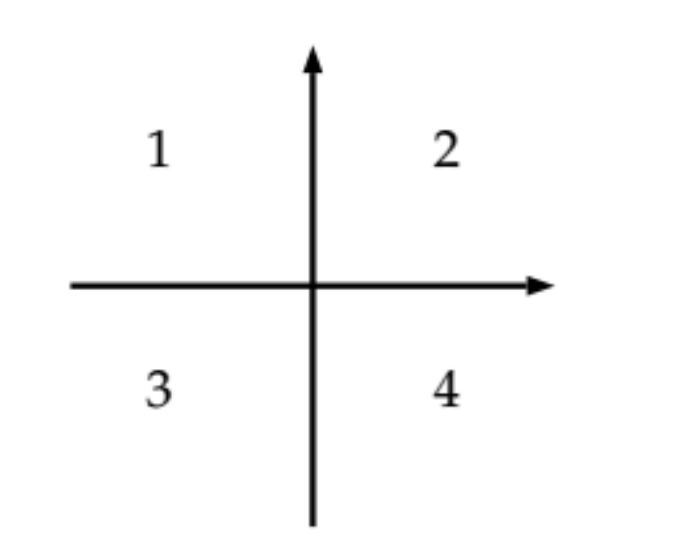}
\caption{Focal plane tessellation for the 4-sided Pyramid WFS. The 1, 2, 3 and 4 indices correspond the the mask's numbers.\label{4uu}}
\end{figure}

In this appendix we apply the previous developments to the most famous and used \cite{Espo03, Espo14, MagAO, Subaru} Fourier-based WFS which is the Pyramid Wave Front Sensor. We study the following configuration: a 4-sided pyramid with an apex angle large enough to completely separate the pupil images that this optical device produces. In order to handle independently these 4 pupil images, we assume that they are generated by 4 Fourier based WFSs using 4 different filtering masks $m_1$, $m_2$, $m_3$ and $m_4$. Each of them corresponds to one quadrant of the cartesian tessellation performed by the Pyramid, see Fig. \ref{4uu}. Moreover, we assume that the effective modulation function $\omega$ is identical for these 4 WFSs. 
\begin{eqnarray}
\Delta I_1(\phi) &=& (\mathbb{I}_P\phi) \star \textbf{IR}(m_1,\omega)\\
\Delta I_2(\phi) &=& (\mathbb{I}_P\phi) \star \textbf{IR}(m_2,\omega)\\
\Delta I_3(\phi) &=& (\mathbb{I}_P\phi) \star \textbf{IR}(m_3,\omega)\\
\Delta I_4(\phi) &=& (\mathbb{I}_P\phi) \star \textbf{IR}(m_4,\omega)
\end{eqnarray}
where
\begin{equation}
\textbf{IR}(m_i,\omega) =  2\textbf{Im}\left[\bar{\hat{m_i}}(\hat{m_i} \star \hat{\omega})\right]
\end{equation}
The set of the differential intensities $\{\Delta I_i\}$ may constitute the Pyramid WFS's output. Nevertheless historically we prefer to combine them to define two new signals called the \emph{slopes maps}:
\begin{eqnarray}
S_x&=&(\Delta I_4 + \Delta I_2)- (\Delta I_1+\Delta I_3)\label{Sxx}\\
S_y&=&(\Delta I_1 + \Delta I_2)- (\Delta I_4+\Delta I_3)\label{Syy}
\end{eqnarray}
The slopes maps have many advantages. They firstly improve the linearity range of the PWFS but they also condense in a smaller signal the phase information. Finally they allow to understand physically how the Pyramid performs the wave front sensing: as a matter of fact, $S_x$ (resp. $S_y$) may be seen as the phase derivative in the spatial frequencies space along the $x$-axis (resp. $a$-axis).  Since  \eqref{Sxx} and \eqref{Syy} transform linearly the differential intensities, it is still possible to associate to them two impulses responses. Indeed, if we define
\begin{equation}
\textbf{IR}_x \equiv \big(\textbf{IR}(m_2,\omega)+\textbf{IR}(m_4,\omega)\big)-\big(\textbf{IR}(m_1,\omega)+\textbf{IR}(m_3,\omega)\big)
\end{equation}
\begin{equation}
\textbf{IR}_y \equiv\big(\textbf{IR}(m_1,\omega)+\textbf{IR}(m_2,\omega)\big)-\big(\textbf{IR}(m_4,\omega)+\textbf{IR}(m_3,\omega)\big)
\end{equation}
the  WFS's input/output relation remains convolutional:
\begin{eqnarray}
S_x&=&(\mathbb{I}_P\phi) \star \textbf{IR}_x\\
S_y&=&(\mathbb{I}_P\phi) \star \textbf{IR}_y
\end{eqnarray}
\begin{figure}[htbp]
\centering

\begin{minipage}[c]{0.09\linewidth}
\centering 
\begin{large}
$S_x$

\vspace{2.cm}

$S_y$
\end{large}
\end{minipage}~
\begin{minipage}[c]{0.60\linewidth}
\includegraphics[trim = 0.1cm 0cm 0cm 0cm, clip,width=8cm]{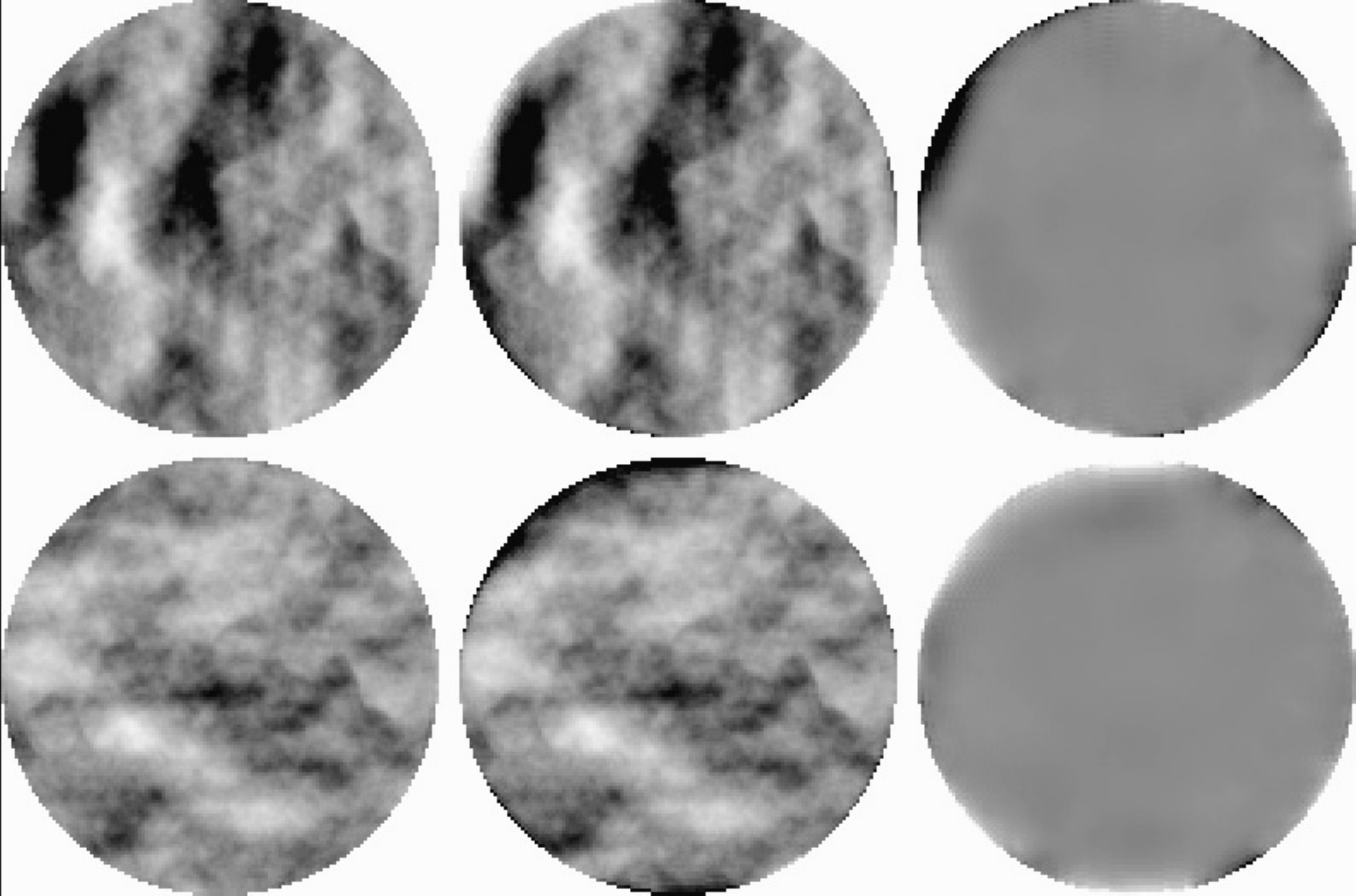}

\small 
~~~~~~~ Linear ~~~~~~~~~ Convolutional  ~~~~~~~~ Residuals

~~~~~~~ model ~~~~~~~~~~~~~~~ model ~~~~~~~~~~~~~~ 
\end{minipage}
\medskip

\caption{Comparison between the slopes maps along $x$ (top) and $y$ (bottom) axis for an arbitrary turbulent phase in the linear (left) and convolutional (middle) models. 
Right inserts give the residuals between the two maps. We assume  a circular pupil and a ring modulation with a modulation radius of 2 $\lambda/D$. The 6 figures have the same greyscale; black corresponds to the mimimum value whereas white corresponds to the maximum.  \label{mIetCie}}
\end{figure}

In order to check the validity of the convolutional model, we give on Fig. \ref{mIetCie} the slopes maps when calculating for the linear (left) and convolutional (middle inserts) models. We choose an input phase which follows a typical atmospheric turbulence. We ensure that the WFS is working in its linearity range. We may observe on the residuals maps (right insert) that these models are in good agreement except at the edge of the pupil. Such a fact confirms the predictions of  \eqref{import} which says that the sliding pupil approximation is not valid on the pupil discontinuities. These results are another proof in favor of phase reconstructors based on deconvolution. 

\begin{figure}[htbp]
\centering
\includegraphics[width=0.65\linewidth]{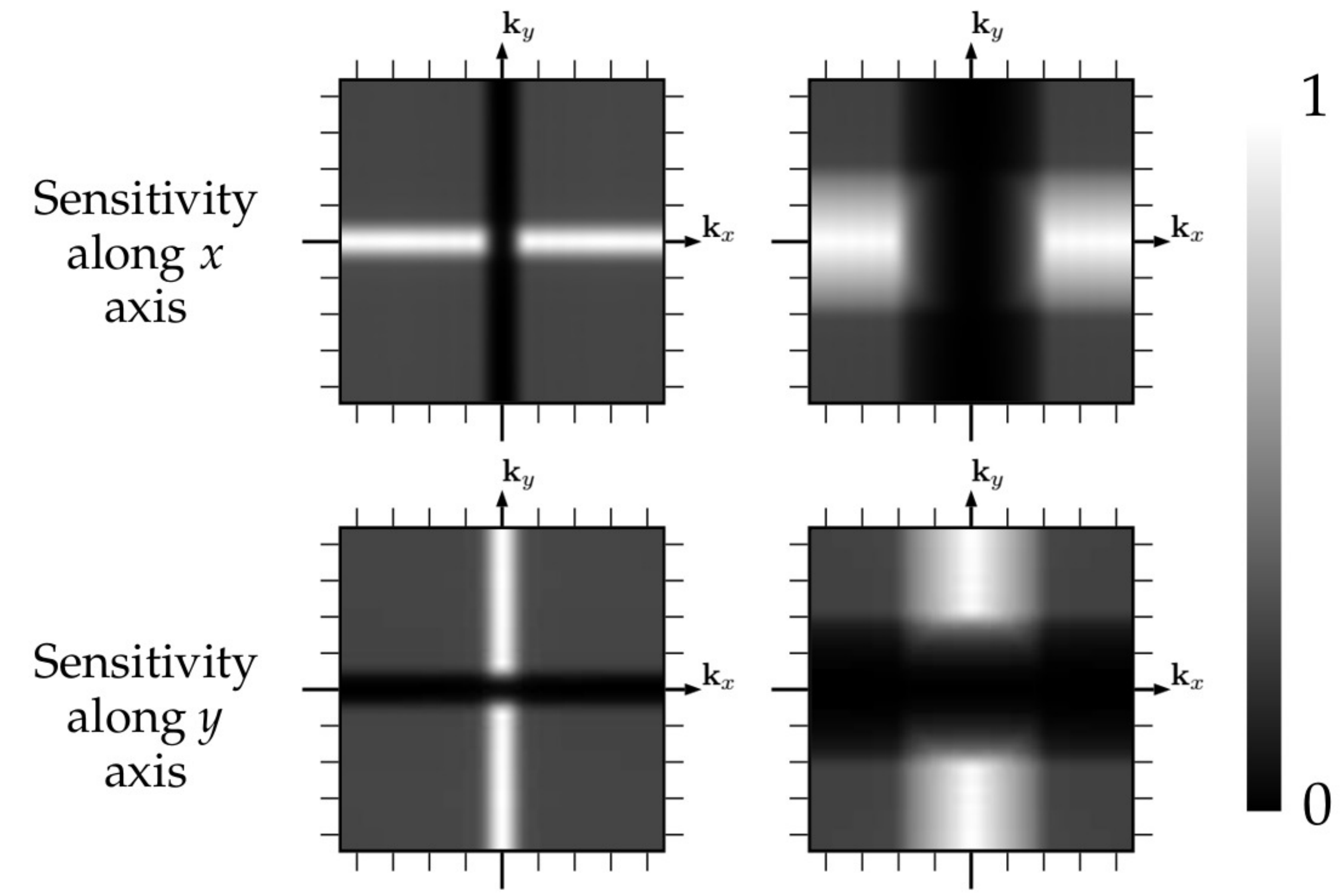}

\caption{Sensitivity with respect to the spatial frequencies associated to the slopes maps along $x$ (top) and $y$ (bottom) axis. Two modulation radii are used: 0.5 (left) and 2 (right) $\lambda/D$. Axis are graduated in $\lambda/D$.\label{ooo}}
\end{figure}

We are now interested in the PWFS's sensitivity with respect to the spatial frequencies. To do so, we just have to know the transfer functions associated to the slopes maps. These ones may be calculated directly from $\textbf{IR}_x$  and $\textbf{IR}_y$ but it is also possible to give the \textbf{TF}s' dependency with respect to the masks and the effective weighting function thanks to  \eqref{symsym}: 
\begin{equation}
\textbf{TF}_x=2\imath\big[m_{3}\star(m_{2}\omega)-m_{2}\star(m_{3}\omega)+  m_{1}\star(m_{4}\omega)-m_{4}\star(m_{1}\omega)\big]
\end{equation}
\begin{equation}
\textbf{TF}_y=2\imath\big[m_{3}\star(m_{2}\omega)-m_{2}\star(m_{3}\omega)-  m_{1}\star(m_{4}\omega)-m_{4}\star(m_{1}\omega)\big]
\end{equation}
It is then possible to use  \eqref{utje} to compute from $\textbf{TF}_x$ and $\textbf{TF}_y$ the sensitivities associated to $S_x$ and $S_y$ with respect to the spatial frequencies: 
\begin{eqnarray}
\text{Sensitivity along $x$ axis} &=&\sqrt{|\textbf{TF}_x|^2 \star  \textbf{PSF}}
\\
\text{Sensitivity along $y$ axis} &=&\sqrt{|\textbf{TF}_y|^2 \star  \textbf{PSF}}
\end{eqnarray}
We give on Fig. \ref{ooo} these sensitivities for two modulation radii: 0.5 $\lambda/D$ on the left inserts and 2 $\lambda/D$ on the right ones. Color scale goes from 0-black which corresponds to non-seen frequencies to 1-white which codes the best seen frequencies. 
For the slopes map $S_x$, we observe that the best sensitivity lies around the $x$-axis while the worst is around the $y$ one. In another word, $S_x$ does measure spatial frequencies in the $x$-direction but is blind to those in the $y$-direction. Fortunately, the other slopes map $S_y$ is complementary. The association of the two slopes maps is therefore able to all the spatial frequencies. Nevertheless, it is worth noticing that low spatial frequencies, i.e. those with are in the midst of the images, have degraded sensitivity in both slopes maps. The number of these badly seen frequencies is even increasing with the modulation radius. We also observe that the transition frequency between this degraded response and the flat-gray response exactly corresponds to the modulation radius. Such results are in perfect agreement with previous works \cite{verinaud2004} about the link between sensitivity and modulation. 

It is also worth mentioning that the total sensitivity is maximum on the neighborhood of the edges of the pyramid mask. Such a fact is not surprising since we know from Foucault and his knife-edge test \cite{Wilson1975} that wave front sensing efficiently works where discontinuities are. We note that this area of maximum signal grows with the modulation radius. It comes from the fact that the modulated Point Spread Function enlarges with the modulation and thus communicates more with the edges of the pyramid mask.

The convolutional model is therefore able to reproduce well-known behaviors of the Pyramid WFS but it also constitutes a significiant improvement in the theoretical understanding of this widely used sensor. As a matter of fact, it is the only 2D model able to accurately mimic the cross structure which as been observed experimentally \cite{bond2018}.

\section{From Integral transform to Matrix transform}\label{tata}

We give in this appendix some elements allowing to understand the link between kernels and matrices. First, we use the lexicographic order for all the 2D images: phase, mask, weighting function, differential intensity, etc. handled in this article. As a consequence, 2D variables become 1D variable:
\begin{eqnarray}
\vec{\textbf{R}} &\rightarrow& X\\
\vec{\textbf{r}} &\rightarrow& x\\
\end{eqnarray}
With such notations, an integral transform may be written:
\begin{equation}
\text{Output}|_X = \int~\text{d}x~\textbf{K}|_{X;x}~\text{Input}|_{x} 
\end{equation}
In order to make appear a matrix transform, we discretize the spatial variables. Previous integral thus becomes a sum:
\begin{equation}
\text{Output}_i = \sum_j~\textbf{K}_{i,j}~\text{Input}_j
\end{equation}
We obviously identify a matrix operation: 
\begin{equation}
\text{Output}=\textbf{K}~\text{Input}
\end{equation}
where \textbf{K} may be understood as:
\newcommand{\mymatrix}[1]{\ensuremath{\left\downarrow\vphantom{#1}\right._X\overset{\xrightarrow[\hphantom{#1}]{~~}}{#1}^x}}
\begin{small}
\begin{equation}
\begin{matrix}
{\begin{pmatrix}
       \text{Output}_{1} \\
        \text{Output}_{2}  \\
        \vdots    \\
        \text{Output}_{n} 
        \end{pmatrix}}&{ =}&\mymatrix{\begin{pmatrix}
       \textbf{K}_{1,1} & \textbf{K}_{1,2} & \cdots & \textbf{K}_{1,n} \\
        \textbf{K}_{2,1} & \textbf{K}_{2,2} & \cdots & \textbf{K}_{2,n} \\
        \vdots  & \vdots  & \ddots & \vdots  \\
        \textbf{K}_{n,1} & \textbf{K}_{n,2} & \cdots & \textbf{K}_{n,n}
        \end{pmatrix}}&{\begin{pmatrix}
       \text{Input}_{1} \\
        \text{Input}_{2}  \\
        \vdots    \\
        \text{Input}_{n} 
        \end{pmatrix}}
\end{matrix}
\end{equation}
\end{small}
$x$ is thus the input space variable while $X$ is the output space variable.

Thanks to this parallel, we may interpret Kernels in the matrix formalism. A typical decomposition of Kernel is for instance:   
\begin{equation}
\textbf{K}|_{X;x} = \textbf{V}|_X \times  \textbf{H}|_x \times  \textbf{C}|_{X-x} \times  \textbf{U}|_{X;x}
\end{equation}
 \textbf{V}, \textbf{H} and \textbf{C} are 1-variable functions and \textbf{K} and \textbf{U} have 2-variables.  $\times$ is the usual scalar multiplication. In the matrix formalism, previous equation may be written:
\begin{equation}
\textbf{K}_{i,j} = \textbf{V}_i \circ  \textbf{H}_j \circ  \textbf{C}_{i-j} \circ  \textbf{M}_{i,j}
\end{equation}
where $\circ$ is the \emph{Hadamard product} which corresponds to a "coefficient by coefficient" multiplication:
\begin{equation}
(\textbf{A}\circ\textbf{B})_{i,j} = \textbf{A}_{i,j}\times\textbf{B}_{i,j}
\end{equation}
Thanks to this interpretation we may detail the structure of \textbf{V}, \textbf{H} and \textbf{C} matrices (there is nothing to say about the \textbf{U}nspecified matrix \textbf{U}). \textbf{H} is a matrix only depending on the input variable $x$, it is thus an horizontal matrix:
\begin{equation*}
\textbf{H} = \begin{pmatrix}
					h_{1}&h_{2}& \dots & h_{{n}}\\
					h_{1}& h_{2}& & h_{{n}}\\
					\vdots &  & \ddots & \vdots \\
					h_{1}& h_{2}& \dots & h_{n}
				\end{pmatrix}
\end{equation*}
In the same way, \textbf{V} only depends on the output variable $X$. Is is a vertical matrix:
\begin{equation*}
\textbf{V} = \begin{pmatrix}
					v_{1}& v_{1}& \dots & v_{{1}}\\
					v_{2}& v_{2}& & v_{{2}}\\
					\vdots &  & \ddots & \vdots \\
					v_{n}& v_{n}& \dots & v_{n}
				\end{pmatrix}
\end{equation*}
\textbf{C} depends on $X-x$. In the matrix formalism, such a type of matrix is called \textbf{circulant}. It has the following form:
\begin{equation*}
\textbf{C} = \begin{pmatrix}
					c_{0}& c_{1}& \dots & c_{{n-1}}\\
					c_{{n-1}}& c_{0}& & c_{{n-2}}\\
					\vdots &  & \ddots & \vdots \\
					c_{1}& c_{2}& \dots & c_{0}
				\end{pmatrix}
\end{equation*}

\bibliography{biblio} 
\bibliographystyle{spiejour}   % makes bibtex use spiejour.bst

\end{document}